\documentclass[aps,preprint]{revtex4}
\usepackage{graphicx}
\usepackage{array}
\newcolumntype{+}{>{\global\let\currentrowstyle\relax}}
\newcolumntype{^}{>{\currentrowstyle}}

\begin{document}
\title{\Large \bf Gravitational Effects of Rotating Braneworld \\Black Holes\footnote{\normalsize  {\it Dedicated to Sergei Odintsov on the occasion of his fiftieth birthday.}}}
\author{\large Alikram N. Aliev and Pamir Talazan}
\address{Feza G\"ursey Institute, P. K. 6  \c Cengelk\" oy, 34684 Istanbul, Turkey}
\date{\today}

\begin{abstract}

We study the light deflection effect and the relativistic  {\it periastron} and {\it frame-dragging}  precessions for a rotating black hole localized on the brane in the Randall-Sundrum  braneworld scenario. Focusing on  a light ray, which passes through the field of the black hole in its  equatorial plane, we first calculate the deflection angle in the weak field limit. We obtain an analytical formula, involving the related perturbative parameters of the field up to the second order. We then proceed with the numerical calculation of the deflection angle in the strong field limit, when the light ray  passes  at the closest distance of approach to the limiting photon orbit. We show that the  deflection angles for the light ray, winding  maximally rotating Kerr and braneworld black holes in the same direction as  their rotation, become essentially indistinguishable from each other for a specific value of the negative tidal charge.  The same feature occurs in the relativistic precession frequencies at characteristic radii, for which the radial epicyclic frequency of the test particle  motion attains its highest value. Thus, the crucial role in a possible identification of the maximally rotating Kerr and braneworld black holes would play their angular momentum, which in the latter case breaches the Kerr bound in  general relativity.

\end{abstract}


\maketitle

\section{Introduction}

The idea of a braneworld  is a leading endeavor to reconcile the properties of gravity in higher dimensions with those in four dimensions \cite{ADD,AADD, RS1, RS2}. According to this idea, our physical world resides on a four-dimensional slice of a fundamental higher-dimensional spacetime. All matter fields  are supposed to be localized on this slice (3-brane) except gravity. Gravity being dynamics of the spacetime itself affects all spacetime dimensions. Therefore, exploring the  behavior of  gravity in braneworld scenarios one may provide a way of tackling the physical signatures of extra unseen dimensions in our four-dimensional world.  In this respect, black holes  might have played an indispensable role.

The first arguments in favor of this claim appear in Arkani-Hamed-Dimopoulos-Dvali (ADD) braneworld scenario \cite{ADD}, where  the large size of the extra spatial dimensions (compared to the Planck length $ \sim 10^{-33} cm $) renders the scale of quantum gravity to be as lower as TeV-energy scales. This opens up the possibility for the formation of mini black holes at these energy scales. Such black holes would carry the imprints of the extra dimensions and their detection at high energy experiments would be a great triumph for theories of gravity in higher dimensions \cite{gt, cavag}. Another intriguing braneworld scenario with a warped and an infinite extra dimension was proposed by Randall and Sundrum (RS) \cite{RS2}, where a single brane is embedded into a five-dimensional anti-de Sitter space (AdS). In the low energy limit, this scenario supports the properties of four-dimensional Einstein's gravity on the brane \cite{sms1, ae1}. Therefore, one can expect the formation of black holes in this scenario due to the gravitational collapse of matter on the brane.

A complete description of black holes in the braneworld scenarios is a thorny and still open problem. There have been several approaches: The first one was build up  on invoking  the classical higher-dimensional black hole solutions. If a black hole on the brane is small enough compared to the size of the extra dimensions, it would  behave as a ``generic"  higher-dimensional object, equally affected by all  spacetime dimensions. In this case, one can use the Tangherlini \cite{tang}, or the Myers-Perry type \cite{mp} solutions to describe these black holes. However, if the size of the  black hole is much larger than that of  the extra dimensions, it would behave like an  effectively four-dimensional object with some finite part of the horizon, leaking into the bulk space. In one approach \cite{chr}, it was suggested to describe these black holes  by the usual Schwarzschild  solution on the RS brane, which from the five-dimensional point of view,  would look  like  a {\it black string} (see also \cite{nojiri, modgil}). The main drawback of this approach is that the black string solution  suffers from curvature singularities, propagating along the extra dimension. In another approach, the braneworld black holes were described by  specifying (postulating)  spherically symmetric,  or  stationary Kerr-Schild type spacetime metrics on the brane and then solving the  effective gravitational equations on it \cite{dmpr, ae2}. This gives rise to a Reissner-Nordstrom type solution  for static black holes \cite{dmpr} and to a Kerr-Newman type solution for rotating black  holes on the brane \cite{ae2}. These solutions carry  a {\it tidal} charge instead of an electric charge, thereby transmitting the gravitational signature of the extra spacelike dimension into the four-dimensional world on the brane.  Further developments in this direction can be found in \cite{whisker, majumdar,pal, roldao, wang, daeho, harko, stuch1, stuch2, lazlo}.

Modern astronomical observations provide compelling and overwhelming evidence for the existence of black holes in the real universe. Today at least twenty black holes and  twenty black hole candidates are known in X-ray binary stellar system (see a recent review \cite{remi2} and references therein). Furthermore, in the X-ray spectra of some black hole binaries high frequency {\it quasiperiodic oscillations} (QPOs)
with single or twin peaks of  characteristic frequencies have been detected \cite{remi2}. The appearance of such frequencies  are of fundamental importance  as they are determined by strong gravity effects near the black holes.  Recent  measurements \cite{nara} also put  the lower bound on the rotation parameter of a black hole in the X-ray binary GRS 1915 + 105; $  a > 0.98 M $, where $ M $ is the mass of the black hole. Clearly, this bound is at a ``threatening" level for black holes in general relativity, for which the rotation parameter  is limited by the Kerr bound $  a = M $.  In the light of all this, it is perhaps good time to ask  a simple and natural question: {\it Are the observed black holes exact prototypes of those predicted by general relativity}?

Today this question is largely open. Future observations will certainly provide crucial  data for more precise measurements of the observable features of black holes, such as the rotation parameter etc. Of course, it may happen that the observational data confront with the predictions of general relativity. This perspective seems to be very exciting and it greatly stimulates theoretical  studies of the observable properties of black holes beyond general relativity (see  for instance, Ref.\cite{horava}). It is also this perspective that  motivates us to explore in this paper some observable effects of rotating braneworld black holes, for which the Kerr bound  $  a = M  $ of general relativity  is violated for negative values of the tidal charge. Namely, we explore the deflection of light and the relativistic  periastron and frame-dragging  precessions in the field of these black holes.

The paper is organized as follows: In Sec.II  we present the metric for  a rotating black hole localized on the brane in the Randall-Sundrum  scenario and  recall some of its basic properties. In Sec.III using the Hamilton-Jacobi formalism, we describe timelike and null geodesics  of this metric. In Sec.IV  we calculate the deflection angle  of a light ray, passing through the field of the rotating  braneworld black hole  in its  equatorial plane. Here we first consider the weak field limit, focusing on the light ray, for which the impact parameter is large enough. Within this approach we derive  an analytical formula for the deflection angle, which involve the associated perturbative parameters of the field up to the second order. We then proceed with the numerical calculation of the deflection angle in the strong field limit, when the light ray  passes  at the closest distance of approach to the limiting photon orbit. In Sec.V employing the theory of cyclic and  epicyclic motions of test particles in an arbitrary stationary and axially symmetric spacetimes, developed earlier in \cite{ag1,ag2, ag3, ag4}, we present analytical expressions  for the orbital,  radial and vertical epicyclic frequencies in the field of the rotating braneworld  black hole. Using these frequencies, we study the relativistic precession effects, periastron and frame-dragging and discuss their observable signatures. In the Appendix  we present the components of the Christoffel symbols for the metric of the rotating braneworld black hole.

\section{Rotating Braneworld Black Holes}

The exact solution for a rotating black hole localized on the  Randall-Sundrum  3-brane was given in \cite{ae2} within the approach that postulates the form of the induced metric on the brane \cite{dmpr}.  Taking the metric ansatz  on the brane  to be of a stationary and axisymmetric Kerr-Schild  form, the authors of \cite{ae2} managed to solve the  effective gravitational equations on the brane in the Randall-Sundrum scenario \cite{sms1, ae1}. When  the bulk space is empty, these equations  have the form
\begin{equation}
R_{ij}=-E_{ij}\,, ~~~~~~i= 0,\, 1,\,2,\,3\,,
\label{breq1}
\end{equation}
where $ E_{ij} $  is the traceless ``electric part" of the
five-dimensional Weyl tensor
\begin{equation}
E_{ij}={}^{(5)} C_{ABCD}\,n^A \,n^C e^B_i e^D_j \,,~~~~~~A= 0,\, 1,\,2,\,3,\,4\, \label{defab}
\end{equation}
and the associated Hamiltonian constraint equation is given by
\begin{equation}
R=0\,. \label{ham1}
\end{equation}
We recall that in the Randall-Sundrum braneworld scenario the
momentum constraint equation is identically satisfied. Furthermore, the cosmological constant on the brane vanishes due to the fine-tuning condition (see for details \cite{ae2}).

Solving equations  (\ref{breq1}) and (\ref{ham1}) with the Kerr-Schild type metric ansatz  on the brane and then passing to the usual  Boyer-Lindquist coordinates, we obtain that the desired metric for the rotating  braneworld black  hole is given by
\begin{equation}
ds^2  =  -{{\Delta}\over {\Sigma}} \left(dt - a \sin^2\theta\,
 d\phi\,\right)^2 + \Sigma \left(\frac{dr^2}{\Delta} + d\theta^{2}\right) + \frac{\sin^2\theta}{\Sigma}
\left[a dt - (r^2+a^2) \,d\phi \right]^2 \,,
\label{ktmetric}
\end{equation}
where the metric functions
\begin{eqnarray}
\Delta &= & r^2 + a^2 -2 M r+ \beta\,,~~~~~~\Sigma=r^2+a^2 \cos^2\theta \,
\label{kmfunc}
\end{eqnarray}
and $ M $ is  the mass, $ a $ is the rotation parameter, or the angular momentum per unit mass, $ a=J/M $, and $ \beta $ is the tidal charge of the black hole.

It should be noted that the field equations (\ref{breq1}) are not closed on the brane  as they involve the quantity $ E_{ij} $ of higher-dimensional origin. Therefore, the metric (\ref{ktmetric}) exactly solves the constraint equation (\ref{ham1}) and when substituting into equation (\ref{breq1}) it closes the system by specifying the ``source" $ E_{ij} $ on the right-hand side. This gives rise to a tidal charge in the metric that creates a Coulomb-type  effect just as the square of the electric charge in the Kerr-Newman  solution. That is, though we have no electric charge on the brane, the rotating black hole solution localized on the brane turns out to be inevitably ``charged"  due to the tidal influence of the bulk space. It is important to note that, unlike the case of the Kerr-Newman solution, the tidal charge in (\ref{ktmetric}) may have  both {\it positive} and {\it negative} values. For $ a=0 $, the metric (\ref{ktmetric}) reduces to a Reissner-Nordstrom type solution, which describes a static braneworld black hole \cite{dmpr}. For $ \beta=0 $, we have the usual Kerr solution.

The physical properties of the metric (\ref{ktmetric}) are largely similar to those of the Kerr-Newman metric in general relativity. However, some significant differences do exist as well. The event horizon structure of the metric (\ref{ktmetric}) is determined by the equation $\,\Delta=0\,$, or equivalently by
\begin{eqnarray}
\Delta &= & r^2 + a^2 -2 M r + \beta =  0\,.
\label{heq}
\end{eqnarray}
The largest root of this equation
\begin{equation}
r_{+}= M + \sqrt{M^2 - a^2- \beta}\, \label{horizon1}
\end{equation}
corresponds to the radius of the horizon. From this equation it follows that  the event horizon exists provided that
\begin{equation}
 M^2 \geq a^2 +\beta\,,
\label{extreme}
\end{equation}
where the equality corresponds to a maximally rotating black hole.  This is the requirement of  a ``Cosmic Censor" (the absence of naked singularities). We note that when the tidal charge is positive, the condition in (\ref{extreme}) gives rise to the  Kerr type bound on the angular momentum. That is, the rotating braneworld black hole with positive tidal charge must possess an angular momentum not exceeding its mass. However, the situation is significantly different for  the negative tidal charge. For a  rotating black hole with $ \beta < 0 $ and $ a=M $, equation (\ref{horizon1}) gives
\begin{equation}
r_{+} = \left(M + \sqrt{ -\beta}\right)\, > M \,.
\label{greaterh}
\end{equation}
This means that for the maximally rotating black hole with the horizon radius $  r_{+} = M  $, the angular momentum is greater than the mass. Thus, {\it braneworld gravity admits a rotating black hole, whose angular momentum exceeds its mass}. This fairly confronts with the Kerr bound in general relativity.

Another important feature of the rotating braneworld black hole is related to the norm of the timelike  Killing vector $ \xi_{(t)}$, which does not vanish on the horizon.  From the equation
\begin{eqnarray}
{\bf \xi}_{(t)} \cdot {\bf \xi}_{(t)}&=& g_{tt}= 0\,,
\label{kproduct}
\end{eqnarray}
we find that
\begin{equation}
r_{0} = M + \sqrt{M^2 - a^2 \cos^2\theta - \beta}\,, \label{ergo0}
\end{equation}
where $ r_{0} $ is the largest root of this equation that describes the boundary of the  ergosphere around the  black hole. It is easy to see from equation (\ref{ergo0}) that the negative tidal charge extends the ergosphere around the braneworld black hole, whereas the positive tidal charge  decreases it \cite{ae2}.  In the limiting case, substituting equation (\ref{extreme}) in equation (\ref{ergo0}), we find that the radius of the ergosphere  falls in the range
\begin{equation}
M<r_0<M+\sin\theta\,\sqrt{M^2-\beta}\,\,. \label{ergob}
\end{equation}
It follows that rotating braneworld black holes with negative tidal charge must be  more energetic objects  compared to those with positive tidal charge.

To conclude this section, we would like to emphasize  once again that the metric (\ref{ktmetric}) was obtained by postulating the form of the induced metric on the brane. This in turn specifies   the form of  the projected electric part $\,E_{ij}\,$ of the bulk space Weyl tensor, thereby closing the system of equations. Of course, the most consistent  approach requires solving the coupled field equations on and off the brane without specifying the form of the quantity $\,E_{ij}\,$. This is still an open problem. Nevertheless, the metric (\ref{ktmetric}) is of particular importance in understanding, at least to some extent, the physical effects of rotating black holes in the braneworld universe.
A linear perturbative analysis shows that the correction to the Newtonian potential  in the RS braneworld scenario behaves  like $\,r^{-3}\,$ \cite{gkr}.  This does  not match the far-distance behavior of the metric (\ref{ktmetric}) that is determined by the $\,r^{-2}\,$ correction term. However, the arguments of \cite{chrss}  show that for a  black hole with mass $ \,M\gg
\ell\,$, where $ \ell\,$ is the curvature radius of AdS space, the post-Newtonian corrections to the metric dominate over the RS-type corrections. That is, a comparison  of the far-distance behavior of the solution (\ref{ktmetric}) with the result of the linear analysis  could be  appropriate only within leading order correction terms.

It should also be emphasized that the negative tidal charge in the braneworld scenario is supposed to be more natural as it causes a  sequestering effect on the brane just as the negative bulk cosmological constant \cite{dmpr} . On the other hand, in \cite{ap1} it is argued  that even the existence of a small positive tidal charge is not supported by modern observations of QPOs in some black hole binaries. Therefore, in the following we basically focus only on the negative value of the tidal charge.

\section{Geodesics}

The description of geodesics in the metric (\ref{ktmetric})  can be carried out using Carter's result \cite{carter1} for the complete separability of  the Hamilton-Jacobi equation in the usual Kerr-Newman spacetime.  For the case of timelike geodesics it was explicitly  done in \cite{ae2}, where substituting the expansion of the action
\begin{equation}
S=\frac{1}{2}m^2\lambda - E t+L \phi  + S_r(r)+S_\theta(\theta)\,,
\label{ss}
\end{equation}
into the Hamilton-Jacobi equation
\begin{equation}
 \frac{\partial S}{\partial \lambda}+\frac{1}{2}\,g^{\mu\nu}\frac{\partial S}{\partial x^\mu} \frac{\partial S}{\partial x^\nu}=0 \,,
 \label{HJeq}
\end{equation}
and focusing on circular geodesics in the equatorial plane $ \theta=\pi/2 $, the authors found that the total energy $ E $ and  the angular momentum $ L $ (along black hole's rotation axis) of a test particle are given by
\begin{equation}
\frac{E}{m}= \frac{r^2-2 M r+\beta \pm a\, \sqrt{Mr-\beta}}
{r\,\left[r^2-3 M r+2 \beta \pm 2\, a \,\sqrt{M
r-\beta}\,\right]^{1/2}}\,\,\,, \label{energy}
\end{equation}
\vspace{3mm}
\begin{equation}
\frac{L}{m}= \pm\,\frac{\sqrt{Mr-\beta}\left(r^2+a^2 \mp 2\,a\,
\sqrt{Mr-\beta}\right) \mp a \beta} {r\,\left[r^2-3 M r+2 \beta \pm
2\, a \,\sqrt{M r-\beta}\,\right]^{1/2}}\,\,\,.\label{momentum}
\end{equation}
Here and in what follows, the upper sign corresponds to direct orbits (the motion of the particle occurs in the same direction as the rotation of the black hole) and  the lower sign refers to retrograde orbits (the particle moves in the opposite direction with respect to the rotation of the black hole). We also note that  the parameter $ \lambda  $ used above is  an  affine parameter along the geodesics and $ m $ is the mass of the test particle. For $ \beta=0 $, the expressions  (\ref{energy}) and (\ref{momentum}) agree with those obtained in \cite{bardeen}.

From equation (\ref{energy}) it follows that the radius of the limiting photon orbit is determined by the equation
\begin{equation}
r^2-3 M r +2 \beta \pm 2\, a \sqrt{M r-\beta}=0 \,.
\label{photont}
\end{equation}
In the general case, when $ \beta \neq 0 $  this equation (\ref{photont}) can be solved only numerically. In particular, one can verify that for $\,a=0\,$ and  $\,\beta=-M^2 \,$,   the radius of the photon orbit $ r_{ph}\simeq 3.56 M $, whereas the radius of the event horizon $\,r_{+}=(1+\sqrt{2}) M \,.$  For the positive tidal charge $\,\beta=M^2 \,$, we have the same limiting radii as for  the Reissner-N\"{o}rdstrom metric. That is,  $ r_{ph}=  2 M $ and  $ r_{+}=M\,$. We recall that for a rotating braneworld black hole with  a negative tidal charge, the rotation parameter $\,a >M \,$. For instance, for $\,\beta=-M^2 \,$ we have the limiting value $\,a=\sqrt{2} M \,$. In this  case,  $\,r_{+}= M \,$ and $ r_{ph} = M  $ in the direct motion, whereas $ r_{ph}\simeq  4.82 M  $  in the retrograde motion. Further details can be found in \cite{ae2}.

For null geodesics $ ( m\rightarrow 0 ) $  the Hamilton-Jacobi equation (\ref{HJeq}) gives the following equations of motion
\begin{eqnarray}
\Sigma\frac{dt}{d\lambda}&=&a(L-aE\sin^2\theta)+
\frac{r^2+a^2}{\Delta}\left[(r^2+a^2)E-aL\right]\,,\label{nullt}\\[3mm]
\Sigma\frac{d\phi}{d\lambda}&=&\left(\frac{L}{\sin^2\theta}
-aE\right)+\frac{a}{\Delta}\left[(r^2+a^2)E-aL\right]\,,\label{nullphi}\\[3mm]
\Sigma\frac{dr}{d\lambda}&=&\sqrt{\mathcal{R}}\,,\label{nullr}\\[3mm]
\Sigma\frac{d\theta}{d\lambda}&=&\sqrt{\Theta}\,,\label{nulltheta}
\end{eqnarray}
where the functions $\mathcal{R}(r)$ and $\Theta(\theta)$ are given by
\begin{eqnarray}
\mathcal{R}&=&\left[(r^2+a^2)E-aL\right]^2-\Delta\left[\mathcal{K}+(L-a E)^2\right]\,,\\[3mm]
\Theta&=&\mathcal{K}+\cos^2\theta\left(a^2E^2-\frac{L^2}{\sin^2\theta}\right)\,,
\end{eqnarray}
and $ \mathcal{K} $ is a constant of separation. These equations describe the propagation of a light ray in  the field of the rotating black hole. Since the spacetime (\ref{ktmetric}) is asymptotically flat, the trajectory of the light ray is a straight line at infinity. In general, the light ray is characterized by two impact parameters, which can be expressed  in terms of the constants of motion and the associated initial  parameters. Using  equations (\ref{nullt}), (\ref{nullphi}) and (\ref{nulltheta}) it is easy to define the impact parameters in the orthogonal and parallel  directions with respect to the axis of  symmetry of the black hole. We have
\begin{eqnarray}
u_{\perp}&=& r^2\sin\theta_0\left(\frac{d\phi}{dt}\right)_{r\rightarrow\infty}
=\frac{L}{E\sin\theta_0}\,,\\[2mm]
u_{\parallel}&=&h\sin\theta_0=
r^2\left(\frac{d\theta}{dt}\right)_{r\rightarrow\infty}=
\frac{\mathcal{K}}{E^2}+\left(a^2-{u_{\perp}}^2\right)\cos^2\theta_0\,,
\end{eqnarray}
where the vertical angle $ \theta_0= \pi/2-\psi_0 $ involves  the inclination angle $ \psi_0 $ between the light ray and the equatorial plane, whereas $ h $ is  the height of the light ray on the equatorial plane.

\section{The Deflection of Light}

The light deflection effect is one of the important predictions of general relativity. It greatly stimulated the famous 1919 eclipse expedition, which  succeeded to confirm general relativity in the weak field, measuring  the deflection of light by the Sun. Clearly, this effect would result in the phenomenon of gravitational lensing, which nowadays has widely been explored in modern observations of galaxies and galaxy clusters (see \cite{schneider} and references therein). In recent developments, the lensing phenomenon has also been studied in the strong gravity regime. The authors of work \cite{ellis} showed that due to a strong deflection  of light near a Schwarzschild black hole, the light ray may several times wind around the black hole, producing the set of observable relativistic images. The light deflection in the strong field regime near the Schwarzschild  black hole was also studied in \cite{bozza1}, where an approximate expression  for the deflection angle in the neighborhood of the limiting photon orbit was obtained. The similar approach was employed to calculate the deflection angle in the strong gravitational field of a nonrotating braneworld black hole \cite{whisker} as well as  in the field of Kerr and Kerr-Sen dilation-axion black holes \cite{bozza2, stoytcho}.

In this section we wish to calculate the deflection of light in  the equatorial plane of the rotating braneworld black hole, focusing on both  the weak and strong field regimes. Using the fact that in the equatorial plane $ \theta=\pi/2 $,\, $ \psi_0=0 $,\, $ u_{\parallel}=0 $  and combining equations (\ref{nullphi}) and (\ref{nullr}), we find that the trajectory of the light ray is described by the equation
\begin{equation}
\frac{d\phi}{dr}=\mathfrak{f}(u,r)\,,
\label{traj}
\end{equation}
where the impact parameter $ u=u_{\perp}=L/E $\, and
\begin{equation}
\mathfrak{f}(u,r)=\frac{(\Delta-a^2)u+a(2Mr-\beta)}
{\Delta\sqrt{(r^2+a^2-au)^2-\Delta(u-a)^2}}\,.
\label{f}
\end{equation}

For practical purposes, it is convenient to express the impact parameter $ u $  in  (\ref{f}) in terms of the distance $ r_0 $ of closest approach to the black hole. Solving equation $ dr/d\phi=0 $, we find that the relation between $ r_0 $ and $ u $ is given by
\begin{equation}
u=\frac{a(\beta-2Mr_0)\pm {r_0}^2\sqrt{\Delta_0}}{{r_0}^2-2Mr_0+\beta}\,\,.
\label{impr0}
\end{equation}
We note that the upper sign refers to the  light ray, winding the black hole in the direction of its rotation ({\it direct winding}), while the lower sign  corresponds to the  case of opposite winding ({\it retrograde winding}). Below, we focus on the direct winding of the light ray. The total shift  in the azimuthal angle  $ \phi $ as the light ray starting  from infinity approaches the  minimal impact distance  $ r_0 $ and then again goes back to infinity is equal to twice of the integral of the  expression in (\ref{traj}) taken from  $ r_0 $ to infinity. Substituting (\ref{impr0}) in (\ref{f}) and taking into account that the  trajectory of  light at infinity is a straight line, we find that the deflection  angle from the straight line is given by
\begin{equation}
\delta\phi=2\int_{r_0}^{\infty}\mathfrak{f}(r,r_0)dr\,\,-\pi\,.
\label{bendingangle}
\end{equation}
Next, we  assume that the light ray passes at a large enough distance from the black hole, $  r_0\gg r_+ \, $. In this case, one can employ the weak field approximation and evaluate the integral analytically by expanding the integrand in small parameters of the weak field $\epsilon=M/r_0$,\,  $\eta=\beta/r_0^2$\, and  $\delta=a/r_0 $. Restricting ourselves to  second order terms in these parameters, we find the analytical  expression for the deflection angle
\begin{equation}
\delta\phi=\frac{4M}{r_0}\left(1-\frac{a}{r_0}\right)
+\frac{M^2}{4r_0^2}(15\pi-16)-\frac{\pi \beta}{4r_0^2}\left(3-\frac{4a}{r_0}\right)+\frac{M \beta}{2r_0^3}\left(3\pi-28\right)+\frac{57\pi}{64}\frac{\beta^2}{r_0^4}\,\,.
\label{anadefang}
\end{equation}
Similar calculations carried out for retrograde winding of the light ray  result in the same expression for  the deflection angle, but with
$ a\rightarrow - a $. The deflection angle  in the latter case is greater for the negative tidal charge of the black hole. For the vanishing  rotation parameter, $ a=0 $,  the expression (\ref{anadefang}) recovers that obtained recently in \cite{lazlo}, whereas for zero tidal charge, $ \beta=0 $, it goes over into the formula given in \cite{irina}.

In general, the integral in (\ref{bendingangle}) is of an elliptic character and can be evaluated only numerically.  Furthermore, when the lower limit of the integral  approaches the radius of limiting photon orbit, $ r_0=r_{ph} $, the integrand  becomes singular. Using the procedure of \cite{bozza2} one can  obtain an approximate analytical expression  for the deflection  angle in terms  of  an expansion  near the limiting photon orbit. This expansion shows a logarithmic divergence when approaching the photon orbit. However, the coefficients of the expansion are  still needed to be evaluated numerically. Therefore, instead of doing this, we perform  the direct numerical evaluation  of the integral in (\ref{bendingangle}) for a light ray, incoming from infinity and  passing close to the limiting photon orbit around the black hole. The results of the numerical calculations are plotted in Figure 1.
 \begin{figure}[!htmbp]%
\centering%
\includegraphics[width=12.5cm]{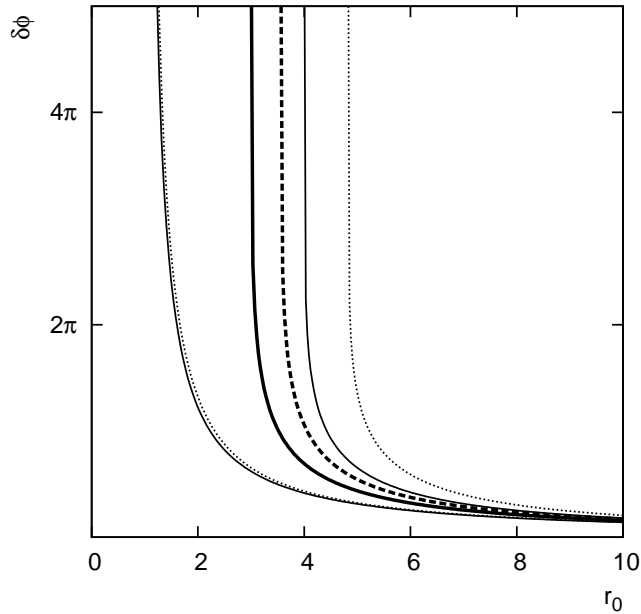}%
\caption{ The deflection angle as a function of  the distance $ r_0 $ of closest approach to the black hole. The pair of curves at the center, the solid and dashed ones, corresponds to a Schwarzschild  black hole and to a static braneworld black hole with the tidal charge $ \beta=-M^2 $, respectively. The pair of curves left to the center, describes  the light deflection for the direct winding the maximally rotating Kerr (the solid one) and braneworld (the dotted one) black holes. Similarly, the pair of curves  right to the center refers to the retrograde case.  In this figure, we have expressed the distances in units of  $ M= 1 $.}%
\label{lens}%
\end{figure}
We see  that with increasing the minimal distance of  approach to the black hole, $ r_0 $, the absolute value of the deflection angle  decreases for both direct and retrograde winding of the light ray. Furthermore, the deflection in the retrograde case is greater and for large enough values of  $ r_0 $ it agrees with the analytical result  (\ref{anadefang}). On the other hand,  as $ r_0 $ decreases, moving towards  the limiting photon orbits, the deflection angle in both direct and retrograde cases increases, reaching at some radii a multiple of  $ 2 \pi $. That is, in the strong gravity regime the photon revolves  several times around the black hole before leaving the region. Finally, when approaching the limiting  radius, $ r_0= r_{ph}\, $, the photon  is captured into infinite revolution and  the deflection angle exhibits the divergence.

The negative tidal charge in general appears to have an increasing effect on the value of the deflection angle compared to the case of zero tidal charge. It is important to note that in the  maximally rotating case and for the direct winding of the light ray, the deflection angles  for  the Kerr black hole $ (a= M ,\, \beta =0) $ and that for the braneworld black hole with the tidal charge $\,\beta=-M^2 $ $ (a=\sqrt{2} M) $  have almost the same value. That is, {\it the light deflection angles for these black holes become essentially indistinguishable from each other} (the pair of curves on the left in Figure 1). Clearly, in this case the angular momentum of these black holes has a crucial meaning, which can be measured from independent observational data. In the retrograde case, the same feature is seen only at distances far enough from the limiting photon orbit, since in this case  the photon capture radii  are different for the Kerr and rotating braneworld black holes (the pair of curves on the right in  Figure 1).  We recall that equation (\ref{photont}) gives $ r_{ph}= 4 M  $  for  the Kerr black hole $ (a= M ) $ and  $ r_{ph}\simeq  4.82 M  $ for the braneworld black hole with $\,\beta=-M^2 $ and  $ a=\sqrt{2} M $.

\section{Epicyclic Frequencies and Relativistic Precessions}

In some cases, especially when studying the equatorial and quasi-equatorial motions of test particles around the black holes, it is convenient to invoke the geodesic equation
\begin{equation}
\frac{d^2 x^{\mu}}{ds^2}+ \Gamma^{\mu}_{\alpha \beta} \frac{d
x^{\alpha}}{ds}\frac{d x^{\beta}}{ds}= 0 \,,
\label{eqmot}
\end{equation}
where the parameter  $ s $  can be thought of as the proper time along the geodesic curves. In particular, to solve the geodesic equation one can  employ the method of successive approximations, which allows us to describe the motion of the test particles in terms of {\it three fundamental frequencies}.  In this approach, the circular motion in the equatorial plane is described at zeroth-order approximation and is characterized by the usual {\it orbital} frequency \cite{bardeen}.
Small perturbations about  circular orbits lead to the quasi-circular ({\it epicyclic})  motion which, at  first-order approximation, amounts to two decoupled  oscillations in the radial and vertical directions. The general theory of the epicyclic motion in the spacetime of stationary black holes was first developed in \cite{ag1} (see also works \cite{ag2, ag3, ag4}). Here, closely following these works, we shall describe the epicyclic motion  in the metric (\ref{ktmetric}). We first begin with recalling some basic ingredient of the theory. Let us introduce a deviation vector
\begin{equation}
\xi^{\mu}(s)= x^{\mu}(s)  - z^{\mu}(s) \,,
\label{expan}
\end{equation}
where
\begin{eqnarray}
z^{\mu}(s) &= & \{t(s) \,\,,r_0\,\, ,\pi/2\,\,,\Omega_0 t(s) \}\,,
 \label{zero1}
\end{eqnarray}
corresponds to the circular motion in the equatorial plane and $ \Omega_0 $ is  the orbital frequency of the motion. The latter is given by
\begin{equation}
\Omega_0 =\frac{\pm\, \Omega_s}{1 \pm a \Omega_s}\,\,,
\label{orbf}
\end{equation}
where we have introduced the  Kepler frequency
\begin{equation}
\Omega_s=\frac{\left(M r - \beta\right)^{1/2}}{r^2}\,.
\label{tkepler}
\end{equation}
Substituting the deviation vector (\ref{expan}) in equation (\ref{eqmot}) and expanding it in powers of  $ \xi^{\mu}(s) $, we obtain the following equation \cite{ag4}
\begin{equation}
\frac{d^2 \xi^{\mu}}{dt^2} + \gamma^{\mu}_{\alpha}\,\frac{d
\xi^{\alpha}}{dt} + \xi^a \partial_a U^{\mu} = {\cal N^{\mu}}\left(\xi, \frac{d
\xi}{dt}\right)\,,~~~~a=1,2\equiv r,\theta \label{perteq}
\end{equation}
where  $ {\cal N^{\mu}} $ stands for all nonlinear terms in  $ \xi^{\mu}(s) $ and we have also used the notations
\begin{eqnarray}
\gamma^{\mu}_{\alpha}&=& 2 \Gamma^{\mu}_{\alpha \beta}
u^{\beta}(u^{0})^{-1} \,,~~~~
U^{\mu}= \frac{1}{2} \, \gamma^{\mu}_{\alpha}\,u^{\alpha}
(u^{0})^{-1} \,. \label{pertquant}
\end{eqnarray}
We recall that all coefficients in this equations, including the quantities   $ \gamma^{\mu}_{\alpha} $ and   $ \partial_a U^{\mu} $  are taken on a circular orbit  $r=r_0,\,\, {\theta=\pi/2}
$\,. At the first-order approximation in  $ \xi^{\mu}(s) $, equation (\ref{perteq}) describes small perturbations around circular orbits. Restricting ourselves to this case and writing down the components of equation (\ref{perteq}), we arrive at the equation for harmonic oscillations in the radial direction
\begin{eqnarray}
\frac{d^2 \xi^{\,r}}{dt^2} +\Omega_{r}^2\, \xi^{\,r}&=& 0\,,
\label{radeq1}
\end{eqnarray}
where the frequency of  radial oscillations is given by
\begin{eqnarray}
\Omega_{r}&=& \left(\frac {\partial U^{r}}{\partial r}-
\gamma_{A}^r\, \gamma_{r}^A
\right)^{1/2}\,,~~~~~~A=0, 3\equiv t,\phi\,.
\label{radfreqs}
\end{eqnarray}
Similarly, writing down equation (\ref{perteq}) for $ \mu=2\, $, we obtain the equation
\begin{eqnarray}
\frac{d^2 \xi^{\,\theta}}{dt^2} +\Omega_{\theta}^2 \,\xi^{\,\theta}
&=& 0 \,,
\label{verteq1}
\end{eqnarray}
which describes harmonic oscillations in the vertical direction with the frequency
\begin{eqnarray}
\Omega_{\theta}= \left(\frac {\partial
U^{\theta}}{\partial \theta}\right)^{1/2}.\label{vertfreq}
\end{eqnarray}
Thus, {\it the epicyclic motion of test particles in a general  stationary and axially symmetric spacetime is in essence equivalent (within the linear approximation) to two decoupled oscillations in the radial and vertical directions} \cite{ag1, ag2, ag3, ag4}.

Next, substituting in the  general expression (\ref{radfreqs})
the components of the Christoffel symbols  given in the Appendix and performing straightforward calculations,  we find that the frequency of radial oscillations is given by
\begin{eqnarray}
\Omega_r^2 &= & \frac{\Omega_0^2}{M r - \beta}\left[M r \left(1-\frac{6M}{r}-\frac{3a^2}{r^2}+\frac{9\beta}{r^2}\right)+\frac{4\beta}{r^2}(a^2-\beta) \pm 8 a \Omega_s(Mr-\beta)\right].
\label{bwradfreq}
\end{eqnarray}
Similarly, using equation (\ref{vertfreq}) we obtain  the frequency of vertical oscillations
\begin{equation}
{\Omega_\theta}^2={\Omega_0}^2\left[1+\frac{a^2}{r^2}
\left(1+\frac{2Mr-\beta}{Mr-\beta}\right)\mp 2a\Omega_s\frac{2Mr-\beta}{Mr-\beta}\right].
\label{bwvertfreq}
\end{equation}
These expressions agree with the uncharged particle limit of the general formulas given in \cite{ag1} for the usual Kerr-Newman metric.
The vanishing of the radial epicyclic frequency, $ \Omega_r^2 =0 $,
determines the radius of the innermost stable circular orbit (ISCO), for which we have the equation
\begin{equation}
Mr\left(1-\frac{6M}{r}-\frac{3a^2}{r^2}+\frac{9\beta}{r^2}\right)+\frac{4\beta}{r^2}(a^2-\beta) \pm 8 a \Omega_s(Mr-\beta) =0\,.
\label{tradstabeq2}
\end{equation}
Solving this equation numerically, we find that for $\,a=0\,$ and $\,\beta=-M^2 \,$, the radius of the ISCO  tends to  $ r_{ms}\simeq 7.3 M  $ and $\,r_{+}=(1+\sqrt{2}) M \,$, whereas  for  $\,\beta= M^2 $,\,\,$ r_{ms}= 4 M $ and $ r_{+}= M $. In the latter case, the radii are the same as those for an extreme Reissner-N\"{o}rdstrom  black hole. On the other hand, for a maximally rotating black hole with the negative tidal charge $\,\beta=-M^2 \,$ and  $\,a=\sqrt{2} M \,$, we obtain that $ r_{ms}= M $ for the direct ISCO and $ r_{ms}\simeq 11.25
M\, $ for the retrograde ISCO. We recall that in this case  $\,r_{+}= M \,$ as well. Full numerical analysis show that the negative tidal charge  has an expelling effect on both direct and retrograde orbits, while the positive tidal charge appears to have the opposite effect \cite{ae2}.

A similar numerical analysis of the expression (\ref{bwvertfreq})
shows that  it is nonnegative in the physical region. That is,
the circular motion around the braneworld black hole is always stable to  linear perturbations in the vertical direction.

Thus, we conclude that one can distinguish  three fundamental frequencies in the field of the rotating braneworld black hole: The frequency of the orbital motion $ \Omega_0 $ and the frequencies $ \Omega_r $ and  $ \Omega_{\theta} $ of the epicyclic motion in the radial and vertical directions, respectively. In the Newtonian regime all these frequencies  coincide with each other, going over into the  Kepler frequency $ (\Omega_0 = \Omega_r = \Omega_{\theta} =  \Omega_s) $. For a static black hole, $ a=0 $, we  have only the equality  $\Omega_0 = \Omega_{\theta} =  \Omega_s $; the frequency of radial oscillations  $ \Omega_r $ is different. However, in the general case  all three frequencies are different from each other. This fact results  in two familiar relativistic precession effects: (i) The effect of  periastron precession, which in the weak field regime  describes the precession of the semimajor axis of elliptic orbits, (ii)  Frame-dragging effect, which in the weak field limit corresponds to the Lense-Thirring precession around a rotating body (see, for instance \cite{chandra}).

The difference between the orbital frequency and the radial epicyclic frequency causes the periastron precession. This leads to a secular shift in the perihelion of an elliptic orbit and the associated angular displacement per one revolution is given by \cite{ag3}
\begin{equation}
\Delta \phi = 2\pi \left|1-\frac{\Omega_0}{\Omega_r}\right|\,.
\label{periangl}
\end{equation}
Consequently, for the frequency of the  periastron precession we obtain
\begin{equation}
\Omega_{PP} = \left|\Omega_0-\Omega_r\right|\,.
\label{periangl}
\end{equation}
Similarly, the failure of the vertical epicyclic frequency to coincide with the orbital frequency of the motion causes dragging of the orbital plane in the direction of rotation. This gravitomagnetic phenomenon is
the reason for the precession of the orbital plane around the axis of symmetry. The precession angle per one revolution  is given by
\begin{equation}
\Delta \phi = 2\pi \left|1-\frac{\Omega_0}{\Omega_{\theta}}\right|\,
\label{lpprec}
\end{equation}
and the corresponding precession frequency is
\begin{equation}
\Omega_{LT} = \left|\Omega_0-\Omega_{\theta}\right|\,.
\label{periangl}
\end{equation}
We recall that as \, $ \Omega_0 > \Omega_{\theta}> \Omega_r $, both precession effects refer to direct orbits. With the radial and vertical epicyclic frequencies given in (\ref{bwradfreq}) and (\ref{bwvertfreq}), it is easy to show that for maximally rotating black holes both  precession frequencies $ \Omega_{PP} $ and $ \Omega_{LT} $ reduce to the angular velocity of the horizon.

It is  important to estimate the value of these frequencies at radii of physical interest. Here we focus on  characteristic radii, at which the radial epicyclic frequency in (\ref{bwradfreq}) reaches its highest value. Passing to the coordinate frequencies $ \nu_{PP} $  and  $ \nu_{LT} $  and expressing them in terms of the characteristic  frequency
\begin{eqnarray}
\nu_l=\frac{\Omega_l}{2\pi}=\frac{c^3}{2\pi G M}\simeq 3.2 \cdot 10^4 \left(\frac{M_{\odot}}{M}\right) Hz\,,
\label{charsc}
\end{eqnarray}
where $ c $ is the speed of light, $ G $ is the  gravitational constant and  $ M_{\odot} $ is the mass of the Sun, we have numerically computed the precession frequencies and their ratio. For the typical mass of a black hole in binary systems  $ M= 10 M_{\odot} $ and for the vanishing tidal charge $ \beta =0 $ (the usual Kerr black hole) the results of calculations are summarized in Table I, whereas for the negative tidal charge $ \beta = - M^2 $ we have  Table II.
\begin{table}[!p]
\renewcommand{\baselinestretch}{1}\normalsize
  \centering
\caption{Relativistic precession frequencies in the Kerr field}
\label{}
\begin{tabular}{|c|c|cc|c|} \hline
\multicolumn{5}{|c|}{~$ \beta=0 \,$,~~$ M=10 M_{\odot}$ }\\
\hline
\rule[-3mm]{0cm}{9mm}
~$a/M$~&~$r_{max}/M$~&$\nu_{PP}$(Hz)&$\nu_{LT}$(Hz)&~$\nu_{LT}/\nu_{PP}$~\\
\hline
0.00 & 8.00 & 70.71 & 0.00 & 0.00 \\
0.10 & 7.58 & 76.89 & 1.43 & 0.02 \\
0.20 & 7.15 & 84.19 & 3.31 & 0.04 \\
0.30 & 6.70 & 92.96 & 5.81 & 0.06 \\
0.40 & 6.24 & 103.71 & 9.24 & 0.09 \\
0.50 & 5.76 & 117.26 & 14.06 & 0.12 \\
0.60 & 5.26 & 134.92 & 21.13 & 0.16 \\
0.70 & 4.71 & 159.16 & 32.12 & 0.20 \\
0.80 & 4.12 & 195.16 & 50.90 & 0.26 \\
0.90 & 3.42 & 257.23 & 89.09 & 0.35 \\
1.00 & 2.42 & 426.28 & 221.21& 0.52 \\
\hline
\end{tabular}
\end{table}
\begin{table}[!p]
\renewcommand{\baselinestretch}{1}\normalsize
  \centering
\caption{Relativistic precession frequencies for nonzero tidal charge}
\label{q-1ltpp}
\begin{tabular}{|c|c|cc|c|} \hline
\multicolumn{5}{|c|}{~$ \beta=-M^2 \,$,~~$ M=10 M_{\odot}$ }\\
\hline
\rule[-3mm]{0cm}{9mm}
~$a/M$~&~$r_{max}/M$~&$\nu_{PP}$(Hz)&$\nu_{LT}$(Hz)&~$\nu_{LT}/\nu_{PP}$~\\
\hline
0.00 & 9.64 & 56.09 & 0.00 & 0.00 \\
0.14 & 9.09 & 61.59 & 1.23 & 0.02 \\
0.28 & 8.53 & 68.17 & 2.87 & 0.04 \\
0.42 & 7.95 & 76.19 & 5.10 & 0.07 \\
0.57 & 7.36 & 86.19 & 8.21 & 0.10 \\
0.71 & 6.74 & 98.99 & 12.68 & 0.13 \\
0.85 & 6.09 & 116.02 & 19.38 & 0.17 \\
0.99 & 5.40 & 139.93 & 30.06 & 0.21 \\
1.13 & 4.64 & 176.41 & 48.83 & 0.28 \\
1.27 & 3.77 & 241.31 & 88.26 & 0.37 \\
1.41 & 2.53 & 426.26 & 231.21& 0.54 \\
\hline
\end{tabular}
\end{table}
Comparing the results displayed in these tables, we have also plotted both cases in Figure 2.
\begin{figure}[!htmbp]%
\centering%
\includegraphics[width=11.9cm]{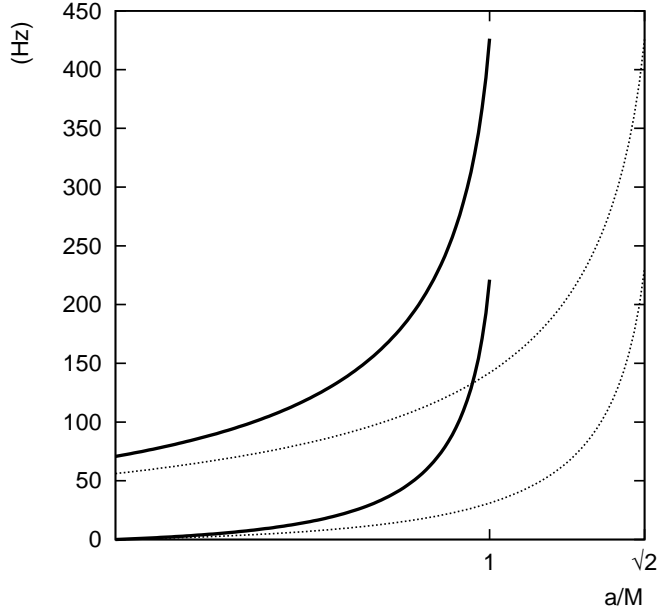}%
\caption{Relativistic precession frequencies as functions of the rotation parameter. The upper solid curve corresponds to $ \nu_{PP} $  and the lower solid curve refers to  $ \nu_{LT} $. Similarly, the dotted curves correspond to the case with nonzero tidal charge.}%
\label{freq-rad-bwbh}%
\end{figure}
This figure clearly shows that for maximally rotating black holes, when the rotation parameter is either $ a= M $ or $ a=\sqrt{2} M $, the corresponding precession frequencies in the field of the black holes with zero tidal charge and with the tidal charge $ \beta=-M^2  $ become essentially indistinguishable from each other. It is also worth noting that in both cases the limiting ratio tends to $ \nu_{PP} : \nu_{LT}= 2 : 1  $. We conclude that {\it for sufficiently fast rotation of the Kerr and braneworld black holes, the relativistic precession frequencies at characteristic radii, for which the radial epicyclic frequency attains its maximum value, fall in the expected ranges of high-frequency QPOs in black hole binaries. Furthermore, the model predicts the special frequency ratios:  $ \nu_{PP} : \nu_{LT}= 3 : 1 $ and  $ \nu_{PP} : \nu_{LT}= 2 : 1  $}.  These results show that the precise measurements of the angular momentum from independent observations (for instance, from relativistically-broadened  Fe K$\alpha$ line formed near the ISCO) would play a crucial role in identification of these  black holes with their real prototypes. In other words, with the detected precession frequencies and their special ratios  given above and  a measured value  of the rotation parameter, obeying the inequality $ a > M $, these results would signal in favor of the existence of a braneworld black hole with a negative tidal charge. That is, in favor of the signature of an extra fifth dimension in the real universe.

\section{Conclusions}

In this paper, we have explored the deflection of light rays and the relativistic  periastron and frame-dragging  precessions in the field of rotating braneworld black holes in the Randall-Sundrum  scenario.
These black holes carry a tidal charge, which transmits the gravitational signature of the fifth dimension into our observable world on the brane. For the vanishing tidal charge, the spacetime metric  recovers the usual Kerr spacetime in general relativity. From the theoretical point of view, the value of the tidal charge is not restricted in the braneworld. However, it is true that the tidal charge may have negative values, which in turn result in the violation of the Kerr bound on the angular momentum of the braneworld black holes.

We have calculated the deflection angle of a light ray in the equatorial plane, considering both the weak and strong gravity regimes. We have derived an  analytical formula for the deflection angle up to the  second order in parameters of  the weak field approximation. In the strong gravity regime, we have employed a numerical method to calculate the deflection angle. We have shown that as the impact parameter of the light ray  decreases, moving towards  the limiting photon orbit, the deflection angle increases and at some radii becomes a multiple of  $ 2 \pi $. That is, in the strong gravity regime the light ray  several times winds around the black hole before leaving the region.  It should be noted that for the negative tidal charge, the deflection angle for the light ray, winding the black hole in the direction of its rotation is always less compared to the case of opposite winding. However, the most striking fact is that for the  specific value of the negative tidal charge  $ \beta= - M^2 $,  the  deflection angle for the light ray, winding  the maximally rotating  braneworld black hole $ (a =\sqrt{2} M) $  and that for the corresponding Kerr black hole $ (\beta =0, a = M) $ have almost the same value.

We have also presented analytical expressions  for the orbital,  radial and vertical epicyclic frequencies of the test particle motion around the rotating braneworld  black hole. Using the fact that all these three frequencies fail to coincide with each other and focusing on characteristic radii, for which the radial epicyclic frequency reaches its maximum value, we have managed to perform the full numerical analysis of the periastron and frame-dragging precessions. We have shown that  for sufficiently fast rotation of the braneworld black hole, the relativistic precessions can be thought of as viable effects to produce high-frequency QPOs in black hole systems. We have also shown that just as in the case of light deflection effect, the  maximally rotating braneworld black hole with the negative tidal charge $ \beta= - M^2 $, produces almost the same frequencies for the relativistic precessions  as the usual maximally rotating Kerr black hole. Furthermore, the associated frequencies always appear  in ratio $ \nu_{PP} : \nu_{LT}= 2 : 1  $.

Thus, we conclude that these two black holes are essentially  indistinguishable from each other with respect to both their  light deflection effect and  the relativistic periastron and frame-dragging  precessions. In this sense, the specific  value $ \beta= - M^2 $ can be thought of as a restriction on the negative tidal charge of the rotating braneworld black hole. The only physical parameter that  can be used to distinguish between these two black holes is the angular momentum, which for the maximally rotating braneworld black hole is greater than its mass. All together, our results show that  the existence of a rotating braneworld black hole with negative tidal charge in principle does not confront with modern observations of black holes in astronomy.

\section{Acknowledgment}

The authors thank the Scientific and Technological Research
Council of Turkey (T{\"U}B\.{I}TAK) for partial support under the
Research Project 105T437.

\appendix*

\section{The Christoffel Symbols}
Using the conventional expression
\begin{equation}
\Gamma^\mu_{\alpha\beta}=\frac{1}{2}g^{\mu\lambda}\left(
\frac{\partial g_{\lambda\alpha}}{\partial x^\beta}+\frac{\partial g_{\lambda\beta}}{\partial x^\alpha}-\frac{\partial g_{\alpha\beta}}{\partial x^\lambda}\right),
\label{christoffel}
\end{equation}
we calculate the nonvanishing components  of the Christoffel symbols for the metric (\ref{ktmetric}). They are given by
\begin{eqnarray}
\Gamma^0_{0 1} &=& -\frac{(r^2 + a^2) B}{2 \Delta}\,,~~~~\Gamma^1_{0 0} = -\frac{\Delta B}{2 \Sigma}\,,~~~~\Gamma^0_{0 2} =
-\frac{2 M r - \beta}{2 \Sigma^2}\,a^2 \sin2\theta\nonumber\,,
\end{eqnarray}
\begin{eqnarray}
\Gamma^1_{1 1} &=&\frac{\Delta r - (r - M) \Sigma}{\Delta \Sigma}\,,~~~~\Gamma^0_{1 3} = \frac{a \sin^2\theta}{\Delta} \left[\frac{(r^2 + a^2) B}{2} - \frac{(2 M r - \beta) r}{\Sigma}\right]\,,
\nonumber
\end{eqnarray}
\begin{eqnarray}
\Gamma^1_{2 2} &=& -\frac{\Delta r}{\Sigma}\,,~~~~
\Gamma^1_{1 2} = -\frac{a^2 \sin2\theta}{2 \Sigma}\,,~~~~
\Gamma^0_{3 2} = \frac{2 M r - \beta}{2 \Sigma^2}\,a^3 \sin^2\theta \sin2\theta\,,
\nonumber
\end{eqnarray}
\begin{eqnarray}
\Gamma^1_{3 3} &=& -\frac{\Delta}{2 \Sigma}  \sin^2\theta \left(2 r + B a^2  \sin^2\theta\right)\,,~~~~
\Gamma^2_{0 0} = -\left(\frac{2 M r - \beta}{2 \Sigma^2}\right)\frac{a^2 \sin2\theta}{\Sigma}\,,
\nonumber
\end{eqnarray}
\begin{eqnarray}
\Gamma^1_{0 3} &=& \frac{\Delta B }{2 \Sigma}\,a \sin^2\theta\,,~~~~
\Gamma^2_{1 1} = \frac{a^2 \sin2\theta}{2 \Delta \Sigma}\,,~~~~
\Gamma^2_{1 2} = \frac{r}{\Sigma}\,,~~~~
\Gamma^2_{2 2} = -\frac{a^2 \sin2\theta}{2 \Sigma}\,,
\nonumber
\end{eqnarray}
\begin{eqnarray}
\Gamma^2_{0 3} &=& \left(\frac{2 M r - \beta}{2 \Sigma^2}\right)\frac{r^2 + a^2}{\Sigma}\,a \sin2\theta\,,~~~~
\Gamma^3_{0 1} = -\frac{a B}{2 \Delta}\,,~~~~
\Gamma^3_{0 2} = -\frac{2 M r - \beta}{\Sigma^2}\,a \cot\theta\,,
\nonumber
\end{eqnarray}
\begin{eqnarray}
\Gamma^2_{3 3} = -\frac{\sin2\theta}{2\Sigma}\left[r^2 + a^2 + \frac{2 M r - \beta}{\Sigma} \left(2 + \frac{a^2 \sin^2\theta}{\Sigma}\right)a^2 \sin^2\theta\right]\,,
\nonumber
\end{eqnarray}
\begin{eqnarray}
\Gamma^3_{3 2} &=& \frac{\cot\theta}{\Delta}\left[\left(1 - \frac{2 M r - \beta}{\Sigma}\right)\left(r^2 + a^2 + \frac{2 M r - \beta}{\Sigma}\,a^2 \sin^2\theta\right)\right.\nonumber\\[1mm]
&+&\left.\frac{2 M r - \beta}{\Sigma^2}\,(r^2 + a^2)\,a^2 \sin^2\theta\right]\,,\nonumber
\end{eqnarray}
\begin{eqnarray}
\Gamma^3_{1 3} = \frac{r}{\Delta}\left(1 - \frac{2 M r - \beta}{\Sigma}\right) + \frac{a^2 B \sin^2\theta}{2 \Delta}\,,
\label{christs}
\end{eqnarray}
where
\begin{equation}
B = \frac{2 M}{\Sigma} \left(1 - \frac{2 r^2}{\Sigma}\right) + \frac{2 r \beta}{\Sigma^2}\nonumber\,\,.
\end{equation}

\end{document}